\documentclass[twocolumn,prl,aps,showpacs]{revtex4}
\usepackage{epsfig}
\usepackage[english]{babel}
\usepackage{latexsym}
\usepackage{graphics}
\usepackage{subfigure}

\begin{document}

\title{Quantum Chaos of Bogoliubov Waves for a Bose-Einstein Condensate in Stadium
Billiards}
\author{Chuanwei Zhang$^{1,2}$, Jie Liu$^{3,1}$, Mark G. Raizen$^{1,2}$, and Qian Niu%
$^{1}$}

\begin{abstract}
We investigate the possibility of quantum (or wave) chaos for the Bogoliubov
excitations of a Bose-Einstein condensate in billiards. Because of the mean
field interaction in the condensate, the Bogoliubov excitations are very
different from the single particle excitations in a non-interacting system.
Nevertheless, we predict that the statistical distribution of level spacings
is unchanged by mapping the non-Hermitian Bogoliubov operator to a real
symmetric matrix. We numerically test our prediction by using a phase shift
method for calculating the excitation energies.
\end{abstract}
\address{$^1$ Department of Physics, The University of Texas, Austin, Texas 78712-1081, USA\\
$^2$ Center for Nonlinear Dynamics, The University of Texas, Austin, Texas 78712-1081, USA\\
$^3$ Institute of Applied Physics and Computational Mathematics, P.O.Box 100088, Beijing, P. R. of China}

\maketitle

In recent years, the realization of Bose-Einstein condensation (BEC) of
dilute gases \cite{review} has opened new opportunities for studying
dynamical systems in the presence of many-body interactions. However, most
previous investigations have focused on one dimensional or high dimensional
separable systems and the dynamics of BEC in nonseparable systems with two
or more degrees of freedom have not received much attention \cite{nonint}.

In the linear Schr\"{o}\-dinger equation, systems with two or more degrees
of freedom can be characterized by the statistics of energy levels: the
typical distribution of the spacing of neighboring levels is Poisson or
Gaussian Ensembles for separable or nonseparable systems respectively \cite
{book}. In the limit of short wavelengths (geometric optics) \cite{book3},
classical trajectories emerge from the linear Schr\"{o}\-dinger equation and
the two types of quantum statistics have been linked to different classical
behaviors: Poisson to regular motion, while Gaussian Ensembles to chaotic motion. It is
natural to ask whether these findings for the linear Schr\"{o}\-dinger
equation can be generalized to other types of wave equations \cite{Andrew}.
The Bogoliubov equation \cite{castin} obtained from the linearization about
the ground state of the Gross-Pitaveskii (G-P) equation has a purely real
spectrum, and there is also a classical limit in the sense of geometric
optics. It therefore makes sense and will be very interesting to explore the
relationship between the Bogoliubov level statistics and regularity of the
corresponding classical trajectories.

There is, however, an important difference between the two types of
equations; while the Schr\"{o}\-dinger equation is Hermitian, the Bogoliubov
equation is non-Hermitian and its statistics can not readily be predicted by
standard random matrix theory. In fact, the Bogoliubov equation belongs to
the category of symplectic problems, describing linearized motion about
stationary states in nonlinear classical Hamiltonian systems. This can be
easily understood by noting that the G-P equation does have a classical
Hamiltonian structure (of infinite dimensions, though) \cite{liu} and that
the Bogoliubov equation describes excitations about a stationary solution of
the G-P equation. The non-Hermiticity of the Bogoliubov equation makes it
allowable to have complex eigenvalues in general, which signifies
instability of the stationary solution. This will not happen about the
ground state (lowest energy state) which is always stable. Therefore, our
investigation of the Bogoliubov problem should shed light on the behaviors
of motions around stable stationary states in extensive classical
Hamiltonian systems.

In this Letter, we investigate the level statistics of Bogoliubov elementary
excitation in separable circular as well as nonseparable stadium billiards.
These are the excitations of a system of interacting particles in contrast
with the modes of non-interacting particles described by the linear
Schr\"{o}dinger equation \cite{McDonald,Graf}. The classical trajectories of
Bogoliubov waves are found to be regular in circular billiards and chaotic
in stadium billiards. By mapping the non-Hermitian Bogoliubov operator to a
real symmetric matrix, we find the mean field interactions in the condensate
do not change the level statistics of Bogoliubov excitations. This
surprising result is tested numerically by using a phase shift method for
calculating the excitation energy. In the regime of strong interaction and
low excitation energy (phonon), we map the Bogoliubov equation to an
equivalent Schr\"{o}\-dinger equation with Neumann boundary condition and
show that the statistics of Bogoliubov levels are the same as that for the
Schr\"{o}\-dinger equation, although interactions in the condensate do
change the average number of levels up to a certain energy.

Consider condensed atoms confined in a quarter-stadium shaped trap of area $A
$ (with length of the top straight side $L$, radius of the semicircle $R$)
and height $d$, where $d<<R$ so that lateral motion is negligible and the
system is essentially two dimensional \cite{ketterle}. With only a quarter
of a stadium, one is restricted to a single symmetry class of the full
problem \cite{McDonald}. The dynamics of the BEC are described by the G-P
equation 
\begin{equation}
i\frac \partial {\partial t}\psi =-\frac 12\bigtriangledown ^2\psi +gN\left|
\psi \right| ^2\psi ,  \label{1}
\end{equation}
where $g=2\sqrt{2}a/d$ is the scaled strength of nonlinear interaction, $N$
is the number of atoms, $a$ is the $s$-wave scattering length. The ground
state of BEC can be written as $\psi =\psi _0\left( \vec{r}\right) \exp
\left( -i\mu t\right) $, where $\mu $ is the chemical potential and $\psi
_0\left( \vec{r}\right) $ can be taken as real. The length and the energy
are measured in units of $b=\sqrt{4A/\pi }$ and $\hbar ^2/mb^2$
respectively, so that the scaled area of billiards is $\tilde{A}=\pi /4$.
The dynamics of the elementary excitations are obtained by linearizing G-P
equation about the ground state $\psi $ and their energy spectrum is
described by the time-independent Bogoliubov equation \cite{castin} 
\begin{equation}
\mathcal{L}\left( \! 
\begin{tabular}{c}
$u$ \\ 
$v$%
\end{tabular}
\ \!\right) =E\left( \! 
\begin{tabular}{c}
$u$ \\ 
$v$%
\end{tabular}
\ \!\right) ,\mathcal{L}=\sigma _z\left( \! 
\begin{tabular}{cc}
$H_1$ \space & $H_2$ \\ 
$H_2$ \space & $H_1$%
\end{tabular}
\ \!\right),  \label{2}
\end{equation}
where $\sigma _z$ is the Pauli matrix, $H_1=-\frac 12\bigtriangledown
^2+2gN\psi _0^2-\mu $, $H_2=gN\psi _0^2$, $E$ is the Bogoliubov excitation
energy, and $(u,v)$ is the eigenwavefunction of linear operator $\mathcal{L}$%
. The ground state wavefunction $\psi _0$ (excitation $\left( u,v\right) $)
has the normalization $\int_{\tilde{A}}\psi _0^2dxdy=1$ ($\int_{\tilde{A}%
}(u^2-v^2)dxdy=1$) and satisfies the boundary condition $\psi _{0\partial 
\tilde{A}}=0$ ($(u,v)_{\partial \tilde{A}}=0$).

%%%%%%%%%%%%%%%%%%%%%%%%%%%%%%%%%%%%%%%%%%%%%%%%%%%%%
\begin{figure}[t]
\begin{center}
\vspace*{-0.5cm}
\par
\resizebox *{6cm}{4cm}{\includegraphics*{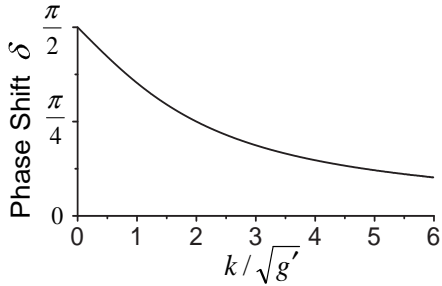}}
\end{center}
\par
\vspace*{-0.5cm}
\caption{ Plot of phase shift $\delta $ versus $k/\protect\sqrt{g^{\prime}}$%
. }
\label{fig:phase2}
\end{figure}
%%%%%%%%%%%%%%%%%%%%%%%%%%%%%%%%%%%%%%%%%%%%%%%%%%

Classical trajectories or rays arise from the Bogoliubov equation in
geometric optics approximation \cite{book3,nonint}. Assume a trial
Bogoliubov wave of the form $\left( u,v\right) =\left( \alpha ,\newline
\beta \right) e^{iS}$ and consider a slowly-varying medium ($\nabla \alpha $%
, $\nabla \beta $ small) and a slowly-varying velocity ($\nabla ^2S$ small)
approximation. We obtain the Eikonal equation $\left| \nabla S\right| ^2=p^2$
with $p=\sqrt{\mu +\sqrt{E^2+\left( gN\psi _0^2\right) ^2}-2gN\psi _0^2}$.
The classical trajectories of the Bogoliubov waves are still governed by the
ray equation $\frac d{dw}\left( p\hat{w}\right) =\nabla p$, where $\hat{w}$
is the direction of the trajectory and $w$ is the arc-length coordinate
along the trajectory. Interestingly, the classical trajectories are simply
straight lines for non-interacting as well as interacting uniform gases
because $p=\sqrt{E_k}$, the local momentum in both cases. In the regime of
strong interaction where the ground state of BEC is nearly uniform, the
classical trajectories of Bogoliubov waves are straight lines and undergo
elastic specular reflection law at the boundary of the billiard. Therefore
we predict that the Bogoliubov level statistics are still Poisson in
circular billiards and Gaussian Orthogonal Ensembles (GOE) in stadium
billiards through quantum classical correspondence.

This prediction is supported by a general argument based on mapping the
non-Hermitian Bogoliubov operator $\mathcal{L}$ to a real symmetric matrix.
The linear operator $\mathcal{L}$ can be written as $\mathcal{L}=\sigma
_{z}Q $, where $Q$ is a real symmetric positive definite matrix because the
ground state of BEC is thermodynamical stable \cite{castin}. The positive
definiteness of $Q$ yields the decomposition $Q=T^{\dagger }T$ ($T$ is a
real matrix with nonzero eigenvalues) and the Bogoliubov equation reduces to 
\begin{equation}
T\sigma _{z}T^{\dagger }\left( \!T\left( \! 
\begin{tabular}{c}
$u$ \\ 
$v$%
\end{tabular}
\ \!\right) \!\right) =E\left( \!T\left( \! 
\begin{tabular}{c}
$u$ \\ 
$v$%
\end{tabular}
\ \right) \right) .  \label{3}
\end{equation}
Therefore Bogoliubov excitation energy is the eigenvalue of a real symmetric
matrix $T\sigma _{z}T^{\dagger }$ and should have GOE distribution in
stadium billiards for arbitrary interaction strength \cite{book}.

In the following, we report numerical test of this prediction by developing
a phase shift method to calculate the Bogoliubov excitation energy. Notice
that the condensate density is nearly uniform \cite{oned} in the interior of
billiards that yields the planewave forms of the Bogoliubov excitations.
Similar to the scattering wave method in linear quantum mechanics \cite
{scattering}, the nonuniform condensate density close to the boundary and
the hard walls can be taken as a pseudopotential and the scattering by this
pseudopotential only induces a phase shift of the excited planewave in the
interior of the billiard. The phase shift may be determined by solving the
one-dimensional G-P equation with an infinite wall at $x\leq 0$. Far from
the wall, the Bogoliubov equation has both planewave and exponential
solutions for a certain excitation energy. We numerically integrate \cite
{recipe} the one-dimensional Bogoliubov equation with two different initial
conditions to eliminate the exponential terms and extract the planewave
solutions. The phase shift $\delta $ is obtained by comparing the numerical
solution with the expected $\sin (kx+\delta )$ dependence. The result is
shown in Fig. 1 as a function of $k/\sqrt{g^{\prime }}$, where $g^{\prime
}=gN\varphi _0^2$, and $\varphi _0$ is the condensate wavefunction far from
the wall. We see that the phase shift approaches $\pi /2$ in the limit of
low energy excitation and strong interactions (phonon) and asymptotically
approaches zero in the regime of high excitation energy and weak interaction
(free particles). As expected, the phase shift is always zero for $g^{\prime
}=0$.

%%%%%%%%%%%%%%%%%%%%%%%%%%
\begin{table}[!t]
\vspace*{-0.5cm}
\caption{Comparison of Bogoliubov excitation energies in one dimensional
billiards using both phase shift ($E_1$) and matrix diagonalization methods (%
$E_2$). $gN=1000$.}
\label{tab:par}%
\begin{ruledtabular}
\begin{tabular}{ccccccccccc}
$E_1$  & $E_2$ & \vline & $E_1$ & $E_2$ & \vline & $E_1$ & $E_2$ & \vline & $E_1$ & $E_2$ \\
\hline 
    0    & 2E-12 & \vline & 907.4 & 909.7     & \vline & 2144.4 & 2147.4 &\vline & 3894.3 & 3897.5 \\
105.8 & 102.6 & \vline & 1038.7 & 1041.1  & \vline & 2333.0 & 2336.0 & \vline & 4153.1 & 4156.6 \\
212.2 & 213.6 & \vline & 1175.9 & 1178.3  & \vline & 2529.7 & 2532.8 & \vline & 4421.4 & 4424.9 \\
320.1 & 322.0 & \vline & 1319.6 & 1321.9  & \vline & 2735.3 & 2738.2 & \vline & 4699.1 & 4702.6 \\
430.4 & 432.4 & \vline & 1470.0 & 1472.3  & \vline & 2949.0 & 2952.3 & \vline & 4968.1 & 4989.6 \\
543.6 & 545.6 & \vline & 1627.0 & 1629.7  & \vline & 3171.9 & 3175.1 & \vline & 5282.2 & 5286.0 \\
660.4 & 662.5 & \vline & 1791.7 & 1794.5  & \vline & 3403.6 & 3406.9 & \vline & 5588.4 & 5591.9 \\
781.5 & 783.6 & \vline & 1964.1 & 1967.0  & \vline & 3644.1 & 3647.6 & \vline & 5903.7 & 5907.3 \\

\end{tabular}
\end{ruledtabular}
\vspace*{-0.0cm}

\end{table}
%%%%%%%%%%%%%%%%%%%

%%%%%%%%%%%%%%%%%%%%%%%%%%%%%%%%%%%%%%%%%%%%%%%%%%%%%
\begin{figure}[!b]
\begin{center}
\vspace*{-0.5cm}
\par
\resizebox *{8cm}{6cm}{\includegraphics*{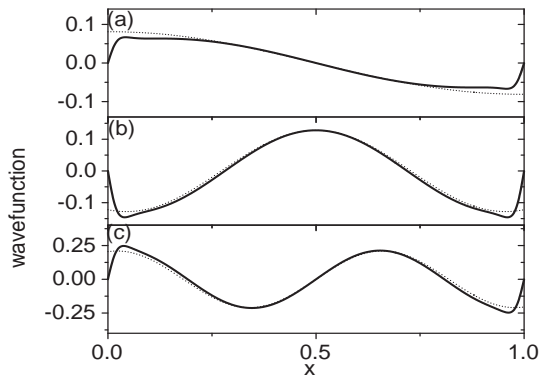}}
\end{center}
\par
\vspace*{-0.5cm}
\caption{Bogoliubov excitation wavefunctions in one dimensional billiards
for $gN=1000$. Solid lines are from the matrix diagonalization method and
dotted lines are from phase shift method $\sin(kx+\delta)$. (a), (b) and (c)
represent the first three excitation wavefunctions, respectively.}
\label{fig:wave}
\end{figure}
%%%%%%%%%%%%%%%%%%%%%%%%%%%%%%%%%%%%%%%%%%%%%%%%%%

With the phase shift method, we can calculate the Bogoliubov excitation
energy in one-dimensional billiards. The reflected planewaves from the
boundary walls $x=0$ and $L$ can be written as $\phi _0=C\sin \left(
kx+\delta _k\right) $ and $\phi _L=F\sin (-k(x-L)+\delta _k)$, respectively.
The continuum of the wavefunction and its derivative in the interior of the
billiards require $\phi _0/\phi _0^{\prime }=\phi _L/\phi _L^{\prime }$ that
yields the quantization condition 
\begin{equation}
kL+2\delta _k=n\pi ,  \label{4}
\end{equation}
where $n$ is an integer. Eq. (4) determines wavevector $k$ and Bogoliubov
excitation energy $E=\sqrt{\frac k2^2\left( \frac{k^2}2+2g^{\prime }\right) }
$ in one-dimensional billiards. For the non-interacting case ($g=0$), the
phase shift $\delta =0$ and Eq. (4) reduces to $k=n\pi /L$, the quantization
condition for a single particle in one-dimensional billiards. As $k$
approaches zero ($E\rightarrow 0$), the phase shift $\delta _k\rightarrow
\pi /2$, therefore $k=0$ ($E=0$) is a solution of Eq. (4) that corresponds
to the ground state of BEC (the uniform density in the interior of the
billiard indicates the $\pi /2$ phase shift).

To check the validity of the phase shift method, we also calculate the
Bogoliubov excitation energies using traditional matrix diagonalization
method in which the linear operator $\mathcal{L}$ is represented as a matrix
and the diagonalization process gives the excitation energy $E$. The results
are compared with those from the phase shift method in Table 1. We see that
the phase shift method gives accurate results for the Bogoliubov excitation
energy. The wavefunction of the first three excited states is shown in Fig.
2. Clearly, the excitation wavefunction is described by the planewave $\sin
\left( kx+\delta \right) $ in the interior of the billiards and drops to
zero at the boundary.

In the regime of weak interaction and high excitation energy ($%
k^2/2>>2g^{\prime }$), the excitations behave like free particles where the
spectrum is well understood. Direct calculations of Bogoliubov excitation
energies in two-dimensional billiards for arbitrary interaction strength and
excitation energy are difficult for both phase shift and matrix
diagonalization methods. However, we are more interested in the regime of
strong interaction and low excitation energy ($k^2/2<<2g^{\prime }$,
phonon), where the effect of interactions is essential. In this regime, the
phase shift $\delta $ is approximately $\pi /2$ as seen from Fig. 1, which
means that the first derivative of the planewave should be zero at the
boundary, instead of the zero wavefunction for the single particle case. The
Bogoliubov wavefunction far from the wall can be written as $\left(
u_k,v_k\right) =\left( U_k,V_k\right) \phi $ , where $\left( U_k,V_k\right)
=\frac 12\left( \chi +\chi ^{-1},\chi -\chi ^{-1}\right) $, $\chi =\left( 
\frac{k^2/2}{k^2/2+2g^{\prime }}\right) ^{1/4}$ \cite{castin}, $\phi $
satisfies the Schr\"{o}\-dinger equation with Neumann boundary condition 
\begin{equation}
\nabla ^2\phi +k^2\phi =0,\text{ }\frac{\partial \phi }{\partial \vec{n}}=0,
\label{5}
\end{equation}
$\vec{n}$ is the normal direction of the billiard wall. Eq. (5) has an
analytical solution $\phi =BJ_{2m}\left( kr\right) \cos \left( 2m\theta
\right) $ in quarter-circular billiards, where $J_{2m}\left( kr\right) $ is
the Bessel function, $m$ is an integer, and $B$ is a normalization constant.
The boundary condition is satisfied by requiring $J_{2m}^{\prime }\left(
kR\right) =0$ that determines wavevector $k$ and Bogoliubov excitation
energy. No analytical solution is available for stadium billiards and we use
the ansatz of the superposition of planewaves $\phi
(x,y)=\sum_{j=1}^Ma_j\cos (k_{jx}x)\cos (k_{jy}y)$ (planewave decomposition
method \cite{heller}) , where $k_{jx}=k\cos (\theta _j),k_{jy}=k\sin (\theta
_j)$, $M$ is the number of planewaves, $\theta _j=2j\pi /M$, i.e. the
direction angles of the wave vectors are chosen equidistantly. The ansatz
solves Eq. (5) inside the billiards and the boundary condition determines
the possible wavevector $k$ and Bogoliubov excitation energy.
%%%%%%%%%%%%%%%%%%%%%%%%%%%%%%%%%%%%%%%%%%%%%%%%%%%%%
\begin{figure}[!b]
\begin{center}
\vspace*{-0.3cm}
\par
\resizebox *{8cm}{4cm}{\includegraphics*{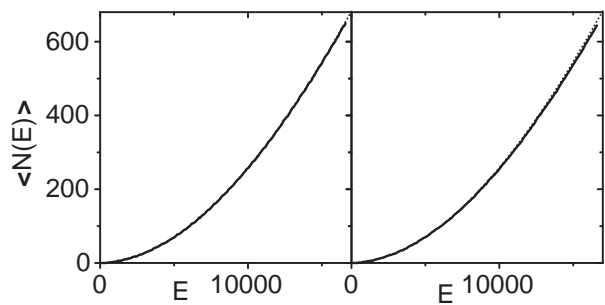}}
\end{center}
\par
\vspace*{-0.5cm}
\caption{Plots of the average number of energy levels $\left\langle N\left(
E\right) \right\rangle $ up to energy $E$ in quarter-circular (a) and
quarter-stadium (b) billiards. $g^{\prime }=gN\varphi_{0}^{2}=25000$. Solid
and dotted lines are from numerical results and Eq. (6), respectively. }
\label{fig:Ne}
\end{figure}
%%%%%%%%%%%%%%%%%%%%%%%%%%%%%%%%%%%%%%%%%%%%%%%%%%

The average number of energy levels $\left\langle N\left( E\right)
\right\rangle $ up to $E$ should satisfy a Weyl-like type formula \cite{weyl}
\begin{equation}
\left\langle N\left( E\right) \right\rangle =\frac{1}{4\pi }\left( \mathcal{A%
}E^{\prime }+\mathcal{D}\sqrt{E^{\prime }}+\mathcal{C}\right),  \label{6}
\end{equation}
where $E^{^{\prime }}=k^{2}=2g^{\prime }\left( \sqrt{1+\left( \frac{E}{%
g^{\prime }}\right) ^{2}}-1\right) $, $\mathcal{A}$ and $\mathcal{D}$ are
the area and the perimeter of the billiard, and $\mathcal{C}$ is a constant
related to the geometry and topology of the billiard boundary. Eq. (6) is
only valid in the semiclassical limit $E^{^{\prime }}=k^{2}\gg \left( \frac{%
\mathcal{D}}{\mathcal{A}}\right) ^{2}$ that yields the Bogoliubov energy $%
\frac{E}{\sqrt{g^{\prime }}}\gg \frac{\mathcal{D}}{\mathcal{A}}\sqrt{1+\frac{%
1}{4g^{\prime }}\left( \frac{\mathcal{D}}{\mathcal{A}}\right) ^{2}}.$ In the
limit of large interaction constant ($E/g^{\prime }<<1$), we have $%
\left\langle N\left( E\right) \right\rangle \approx \frac{1}{4\pi }\left( 
\mathcal{A}\left( \frac{E^{2}}{g^{\prime }}-\frac{E^{4}}{2\left( g^{\prime
}\right) ^{3}}\right) +\mathcal{D}\sqrt{\frac{E^{2}}{g^{\prime }}}+\mathcal{C%
}\right) $with the condition $\frac{\mathcal{D}}{\mathcal{A}}\ll \frac{E}{%
\sqrt{g^{\prime }}}\ll \sqrt{g^{\prime }}$.

%%%%%%%%%%%%%%%%%%%%%%%%%%%%%%%%%%%%%%%%%%%%%%%%%%%%%
\begin{figure}[!t]
\begin{center}
\vspace*{-0.5cm} \resizebox *{8cm}{4cm}{\includegraphics*{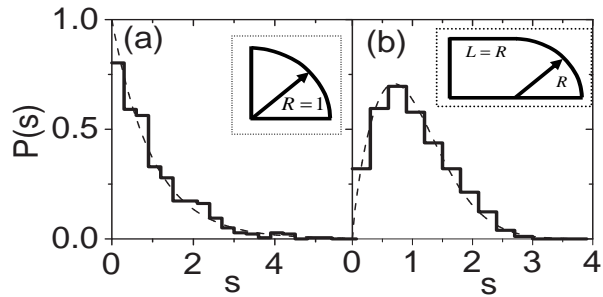}}
\end{center}
\par
\vspace*{-0.5cm}
\caption{The distributions of the spacing of neighboring Bogoliubov energy
levels ( $k<100$ and the lower 50 ones are omitted) in phonon regime. (a)
Circular billiards. Dashed line: Poisson distribution $P_{0}(s)=\exp \left(
-s\right) $. (b) Stadium billiards. Dashed line: Brody distribution $P\left(
s\right) =\left( \nu +1\right) a_{\nu }s^{\nu }\exp \left( -a_{\nu}s^{\nu
+1}\right) $ from the prediction of GOE, where $a_{\nu }=\left[ \Gamma
\left( \frac{\nu +2}{\nu +1}\right) \right] ^{\nu +1}$ and $\Gamma$ is Gamma
function. The fitting Brody parameter $\nu=0.76$ is smaller than expected $%
0.953$ due to the "bouncing ball states" \protect\cite{Graf} and finite
levels effects \protect\cite{Shudo}.}
\label{fig:sta1}
\end{figure}
%%%%%%%%%%%%%%%%%%%%%%%%%%%%%%%%%%%%%%%%%%%%%%%%%%

We have computed the Bogoliubov excitation energies in circular billiards ($%
R=1,L=0$) using the analytical solution and in stadium billiards ($R=L=\sqrt{%
1/(1+4/\pi )})$ using planewave decomposition method. In Fig. 3, we plot $%
\left\langle N\left( E\right) \right\rangle $ in both circular and stadium
billiards. The agreement between the numerical results and the corresponding
Weyl-like formula (Eq. (6)) is satisfactory. We unfold the spectrum formed
by the energies $E_{n}$, i.e., we evaluate Eq. (6) for each $E_{n}$ in order
to obtain the new energies $\tilde{E}_{n}=\left\langle N\left( E_{n}\right)
\right\rangle $. Note that the integer part of $\tilde{E}_{n}$ is about $n$
and, as a result, the corresponding mean level spacing is characterized
through $\left\langle s\right\rangle =\sum \left( \tilde{E}_{n+1}-\tilde{E}%
_{n}\right) /\left\langle N\right\rangle \approx 1$. The resulting level
spacing distributions $P(s)$ are shown in Fig.4. Clearly, the statistics of
Bogoliubov excitation energy levels spacing are still Poisson in circular
billiards and GOE in stadium billiards.

Experimentally, the system can be realized by confining BEC in
two-dimensional optical billiards \cite{billiard}. Atoms could be trapped in
a one dimensional optical lattice that is in the vertical direction to
counteract gravity. Transverse motion could be confined to a closed pattern
of light that is tuned to the blue side of atomic resonance, forming a
repulsive barrier for the atoms. This pattern can be created by rapid
scanning of a beam as was demonstrated for ultracold atoms. However, for the
case of a BEC it would be better to created a static patterns, which can be
accomplished using a liquidcrystal spatial light modulator \cite{Roy}. The
interaction strength may be adjusted by the confinement in the vertical
direction, or by the number of atoms in each node of the standing wave. The
Bogoliubov excitation energy may be measured using Raman transition between
two hyperfine ground states, denoted $\left| 1\right\rangle $ and $\left|
2\right\rangle $ respectively. The condensate would be formed in state $%
\left| 1\right\rangle $. Two co-propagating Raman beams would drive a
transition to state $\left| 2\right\rangle $ and the number of atoms in that
state would be measured as a function of the frequency difference in the
beams. The coupling efficiency to the excited states must depend upon the
spatial profiles of the beams, requiring more detailed analysis \cite{pre}.

We acknowledge the support from the NSF, Quantum Optics Initiative US Navy -
Office of Naval Research, Grant No. N00014-03-1-0639, and the R. A. Welch
foundation, MGR also acknowledges supports from the Sid W. Richardson
Foundation. We thank L.E. Reichl and E.J. Heller for helpful comments.

\end{document}